\documentclass[a4paper,amsmath,amssymb,superscriptaddress,nofootinbib,11pt]{article}
\usepackage{url}
\pdfoutput=1 

\usepackage{jheppub} 

\usepackage[T1]{fontenc} 

\usepackage{epsfig}
\usepackage{amsfonts}
\usepackage{graphicx}
\usepackage{epsfig}
\usepackage{eepic}
\usepackage{amsmath}
\usepackage{amssymb}
\usepackage{color}
\usepackage{bbm}
\usepackage{dcolumn}
\usepackage{bm}
\usepackage{ulem}
\usepackage{mathrsfs}
\usepackage{bbold}
\usepackage{datetime}

\usepackage{overpic} 
\usepackage{rotating}
\usepackage[usenames,dvipsnames]{xcolor}

\def\bc{\begin{center}}

\def\ec{\end{center}}
\def\be{\begin{eqnarray}}
\def\ee{\end{eqnarray}}
\definecolor{dyellow}{rgb}{1.,0.8,.0}
\definecolor{myblue}{rgb}{.1,.1,.7}
\definecolor{dcyan}{rgb}{.0,.6,.6}
\definecolor{dmagenta}{rgb}{0.6,0.0,0.6}
\definecolor{brown}{rgb}{0.6,0.2,0.}
\definecolor{darkblue}{rgb}{.0,.0,0.5}
\definecolor{darkred}{rgb}{0.75,0.0,0.0}
\definecolor{orange}{rgb}{1.,.6,.0}
\definecolor{dorange}{rgb}{0.8,.4,.0}
\definecolor{darkgreen}{rgb}{0.0,0.6,0.0}
\definecolor{purple}{rgb}{.4,.0,.4}
\definecolor{lightgrey}{rgb}{0.7, 0.7, 0.7}
\definecolor{grey}{rgb}{0.4, 0.4, 0.4}

\usepackage[position=t, singlelinecheck=off]{subfig}
\usepackage[font=small,labelfont=bf,justification=raggedright]{caption}

\newcommand{\xdownarrow}[1]{%
  {\left\downarrow\vbox to #1{}\right.\kern-\nulldelimiterspace}
}
\newcommand{\xuparrow}[1]{%
  {\left\uparrow\vbox to #1{}\right.\kern-\nulldelimiterspace}
}

\title{Holographic topological defects in a ring: role of diverse boundary conditions}
\author[a]{Zhi-Hong Li,} 
\author[a]{Han-Qing Shi,}
\author[a,b]{and Hai-Qing Zhang}
\affiliation[a]{Center for Gravitational Physics, Department of Space Science, Beihang University,
Beijing 100191, China}
\affiliation[b]{International Research Institute for Multidisciplinary Science, Beihang University, Beijing 100191, China}

\emailAdd{lizhihong@buaa.edu.cn}
\emailAdd{by2030104@buaa.edu.cn}
\emailAdd{hqzhang@buaa.edu.cn}

\abstract{
We investigate the formation of topological defects in the course of a dynamical phase transition with different boundary conditions in a ring from AdS/CFT correspondence. According to the Kibble-Zurek mechanism, quenching the system across the critical point to symmetry-breaking phase will result in topological defects -- winding numbers -- in a compact ring. By setting two different boundary conditions, i.e., Dirichlet and Neumann boundary conditions for the spatial component of the gauge fields in the AdS boundary, we achieve the holographic superfluid and holographic superconductor models, respectively. In the final equilibrium state, different configurations of the order parameter phases for these two models indicate a persistent superflow in the holographic superfluid, however, the holographic superconductor lacks this superflow due to the existence of local gauge fields. The two-point correlation functions of the order parameter also behave differently. In particular, for holographic superfluid the correlation function is a cosine function depending on the winding number. The correlation function for the holographic superconductor, however, decays rapidly at short distances and vanishes at long distance, due to the random localities of the gauge fields. These results are consistent with our theoretical analysis.}

\begin{document} 
\maketitle
\flushbottom


\section{Introduction}
Topological defects arise in a wide variety of systems ranging from early universe to condensed matter systems \cite{Davis:1990mt, Bunkov:2000yi}. Topological defects mostly form during phase transitions accompanied by spontaneous breaking of discrete or continuous symmetries. In the second order phase transition, formation of topological defects in a dynamical process is described by the Kibble-Zurek mechanism  (KZM) \cite{Kibble:1976sj,Kibble:1980mv,Zurek:1985qw}, which states that as the system temperature approaches the critical temperature $T_c$ from above, the dynamics of the field almost freezes as soon as it enters the regime of critical slowing down and the topological defects will form in this process. The resulting number density of the topological defects has a simple power law to the quench strength with the power containing the static and dynamic critical exponents in the equilibrium state.  The processes of formation and evolution of these defects have been extensively studied in various experiments \cite{Chuang:1991zz,Ruutu:1995qz,Carmi:2000zz} and numerical studies \cite{Laguna:1996pv,Yates:1998kx,Donaire:2004gp}. The reviews \cite{kibblereview,zurekreview} can be consulted.

AdS/CFT correspondence (gauge-gravity duality) \cite{Maldacena:1997re,Witten:1998qj,Gubser:1998bc} is a powerful tool enabling us to study the strongly coupled field theories from the classical gravity in one higher dimensions.  Among these, holographic superfluid/superconductor models is first proposed in \cite{Gubser:2008px,Hartnoll:2008vx}, which involves a charged scalar field living in the bulk of an AdS planar black hole. As the temperature of the black hole is lower enough, the scalar field will develop a nonzero solution near the horizon, which will induce the U(1) symmetry breaking on the boundary, resulting the condensate of the superconducting order parameter due to the Higgs mechanism. Holographic dynamics of U(1) symmetry breaking, or holographic KZM are studied previously in spatial 1D system \cite{Sonner:2014tca,Xia:2020cjl,Xia:2021xap,Li:2021mtd} and 2D system \cite{Chesler:2014gya,Zeng:2019yhi,Li:2019oyz,delCampo:2021rak,Li:2021iph,Li:2021dwp}. 

In the present work, we focus on exploring and comparing the different phenomena, including configurations of phases of the order parameter, the statistics of the winding numbers and the two-point correlation functions in the holographic superfluid and holographic superconductor system in 1D ring. According to the KZM, we dynamically realize the holographic superfluid and holographic superconductor phase transition by imposing the Dirichlet and Neumann boundary conditions for the spatial component of the gauge fields near the AdS boundary, respectively \cite{witten,silva}. The topological defects $-$ winding numbers $-$ are generated during this dynamical phase transition. In the final equilibrium state, the phase configurations of the order parameters of the two models are dramatically different. In particular, in the superfluid model the phase will eventually relax with constant gradient, indicating a persistent superflow along the ring. However, for the superconductor model the phase will finally relax in a random smooth function due to the random locality of the gauge fields.  This in turn implies that the gauge invariant velocity in the holographic superconductor will always remain zero, which is also a requirement from the minimum of free energy. We numerically and statistically verify this statement.  We also compare the two-point correlation functions of the order parameter in the final equilibrium state.  We find that in the superfluid system the correlation function is a cosine function depending on the winding numbers. In particular, if the winding number is zero, the correlation function is a constant indicating perfect correlations in superfluid. Numerical results satisfy the analytical functions very well. However, for the holographic superconductor system the two-point correlation functions show a rapid decay at short distance and vanish at longer distance. We argue that this is due to the randomness of the local gauge fields. 

The plan of this paper is as follows. In Section \ref{setup} we show the holographic setup of this superconducting model; Section \ref{dynamics} studies the dynamical phase transitions and the formation of winding numbers in the holographic superfluid and holographic superconductor models; Section \ref{statistics} is devoted to the statistics of winding numbers and the superflows in the final equilibrium state while Section \ref{2point} is focusing on the two-point functions of the two models; We finally draw the conclusions and discussions in Section \ref{conclusion}.

\section{Holographic setup}
\label{setup}
We adopt the Abelian-Higgs action to study the holographic superfluid/superconductor system \cite{Hartnoll:2008vx},
\begin{equation}\label{density}
S=\int d^4x\sqrt{-g} \left( -\frac{1}{4} F_{\mu \nu} F^{\mu \nu} - |D \Psi|^2 - m^2 |\Psi|^2\right).
\end{equation}
where $F_{\mu\nu}=\partial_\mu A_\nu-\partial_\nu A_\mu$ is the field strength of the U(1) gauge field $A_\mu$ and $D_\mu \Psi=\nabla\Psi -iA_\mu \Psi$, in which $\Psi$ is the complex scalar field $\Psi=|\Psi|e^{i\theta}$. We simply consider the `probe limit' in this paper, hence the equations of motion read
\begin{eqnarray}\label{eomofwhole}
D_\mu D^\mu\Psi-m^2\Psi=0, ~~~~\nabla_\mu F^{\mu\nu}=J^\nu,
\end{eqnarray}
where the current $J^\nu=i\left(\Psi^* D^\nu\Psi-\Psi{(D^\nu\Psi)^*}\right)$.
We use the Eddington-Finkelstein coordinates in the AdS$_4$ black brane to study the dynamics of the system \cite{Chesler:2013lia}, 
\begin{equation}
ds^2 = \frac{1}{z^2} (-f(z) dt^2 - 2dtdz + dx^2+dy^2 ),
\end{equation}
where $f(z) = 1 - (z/z_h)^3$, with $\{z, z_h\}$ representing AdS radial coordinate and the location of horizon respectively (we have scaled the AdS radius $l=1$ and $z_h=1$ for simplicity). The AdS infinite boundary is at $z = 0$ where the field theory lives.  The Hawking temperature associated with the above black brane is given by $T=3/(4\pi)$. Since we are interested in the model consisting of one dimensional ring on the boundary, we can compactify the $x$-spatial direction as the ring and meanwhile ignore the other spatial direction. Therefore, the ansatz for the fields are $\Psi = \Psi(t,z,x), {A_{t} = A_{t}(t,z,x), A_x=A_x(t,z,x)}$ and $A_z =A_y= 0$. {The ansatz of the gauge fields is a well-defined choice for the system, since there are four real independent fields, i.e., complex field $\Psi$, real $A_t$ and $A_x$, to be solved by four independent real equations. Please refer to the Appendix \ref{app} for details.}

{\bf Boundary conditions :}
{Without loss of generality, we set the scalar mass square as $m^2=-2$.} For the scalar field one finds the asymptotic expansion near $z\to0$,
\be
\Psi=z\left(\Psi_0+\Psi_1 z+\dots\right)
\ee
From the AdS/CFT dictionary, $\Psi_0$ is interpreted as the source of a scalar operator ${O}$ on the boundary, while $\Psi_1$ corresponds to the expectation value of this operator $\langle{O}\rangle$. Therefore, we set the source $\Psi_0=0$ in order to satisfy the requirement that the symmetry is spontaneously broken. 
The asymptotic behaviors of gauge fields near $z\to0$ are
\be
 A_\mu\sim a_\mu+b_\mu z+\dots .
\ee 
From holography,  $a_t$ and $b_t$ are interpreted as the chemical potential and charge density on the boundary, respectively.  $a_x$ have different physical meanings on the boundary depending on their boundary conditions \cite{Montull:2009fe}. For instance, if one sets Dirichlet boundary conditions of $a_x$ on the boundary, then $a_x$ can be interpreted as the sample velocities related to a superfluid model; However, if one sets Neumann boundary conditions for $a_x$, then it is related to turning on electromagnetic fields on the boundary for a superconductor model. Indeed, there have been numerous papers to set the Neumann boundary conditions of $a_x$ in order to get the dynamical gauge fields on the boundary which can mimic a genuine superconductor \cite{silva,Zeng:2019yhi,Li:2019oyz,delCampo:2021rak,Li:2021iph,Li:2021dwp}. The sub-leading term $b_x$ is related to the conserved current on the boundary field theory. In addition, we impose the periodic boundary conditions for all the fields along $x$-direction. At the horizon, we demand the regularity of the fields by setting $A_t(z_h)=0$ while other fields are finite.

{\bf Numerical schemes:}
It is known that increasing the charge density in a holographic superfluid/superconductor system is equivalent to lowering the system's temperature \cite{Gubser:2008px,Hartnoll:2008vx}. According to the dimensional analysis, the temperature of the black brane $T$ has mass dimension one, while the charge density $\rho$ on the boundary has mass dimension two. Therefore, $T/\sqrt{\rho}$ is a dimensionless quantity. In order to linearly quench the temperature across the critical point according to KZM as $T(t)/T_c=1-t/\tau_Q$, we quench the charge density $\rho$ as $\rho(t)=\rho_c(1-t/\tau_Q)^{-2}$, where $\tau_Q$ is the quench strength while $\rho_c$ is the critical charge density for the static and homogeneous holographic superconducting system. Before quenching the system, we keep the system at the initial temperature and thermalize it by adding the Gaussian white noise $\xi(x_i,t)$ into the bulk with $\langle \xi(x_i,t)\rangle=0$ and $\langle \xi(x_i,t)\xi(x_j,t')\rangle=h\delta(t-t')\delta(x_i-x_j)$, with a small amplitude $h=0.001$. After this thermalization, we linearly quench the system from $T_i = 1.4T_c$ to $T_f = 0.8T_c$, rendering the system to evolve from a normal state to a superconducting state. The simulation in this paper is performed on a Chebyshev pseudo-spectral grid in the radial direction of $z$ with 21 grid points. Since we are working in a compact ring along $x$-direction, we adopt the Fourier decomposition along $x$ such as ($x \sim x+L$) with $201$ grid points with $L$ the length of the ring. For each run we evolve the system by using the 4th-order Runge-Kutta method with time step $\Delta t=0.1$.

\section{Quenched dynamics \& formation of winding numbers}
\label{dynamics}

\begin{figure}[h]
\centering
\includegraphics[trim=1.cm 11.cm 1cm 7.5cm, clip=true, scale=0.42]{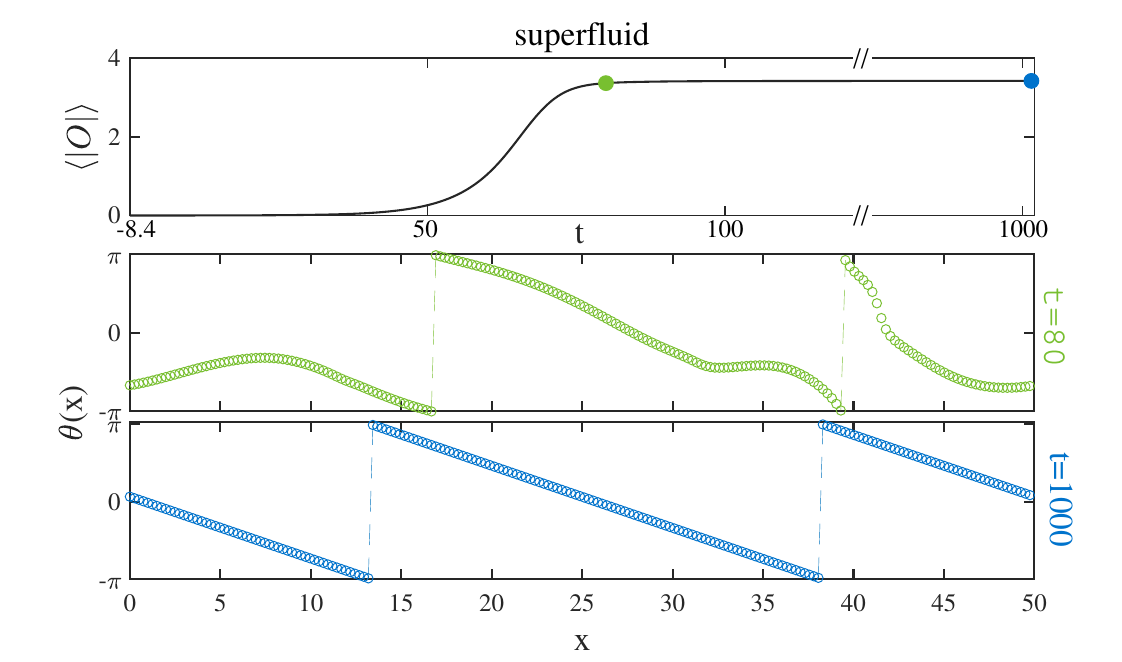}
\put(-220,130){(a)}~
\includegraphics[trim=1.cm 11.cm 1cm 7.5cm, clip=true, scale=0.42]{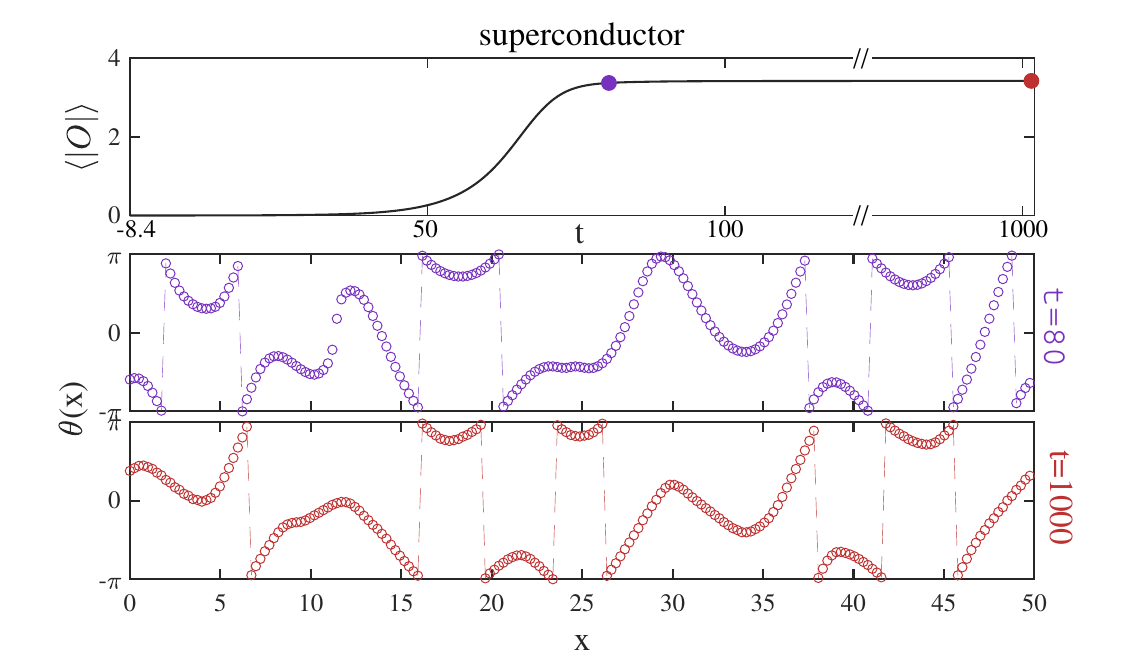}
\put(-220,130){(b)}\\~~~~~~~~~~~~~
\includegraphics[trim=2.4cm 5.3cm 3cm 4cm, clip=true, angle=-90,scale=0.3]{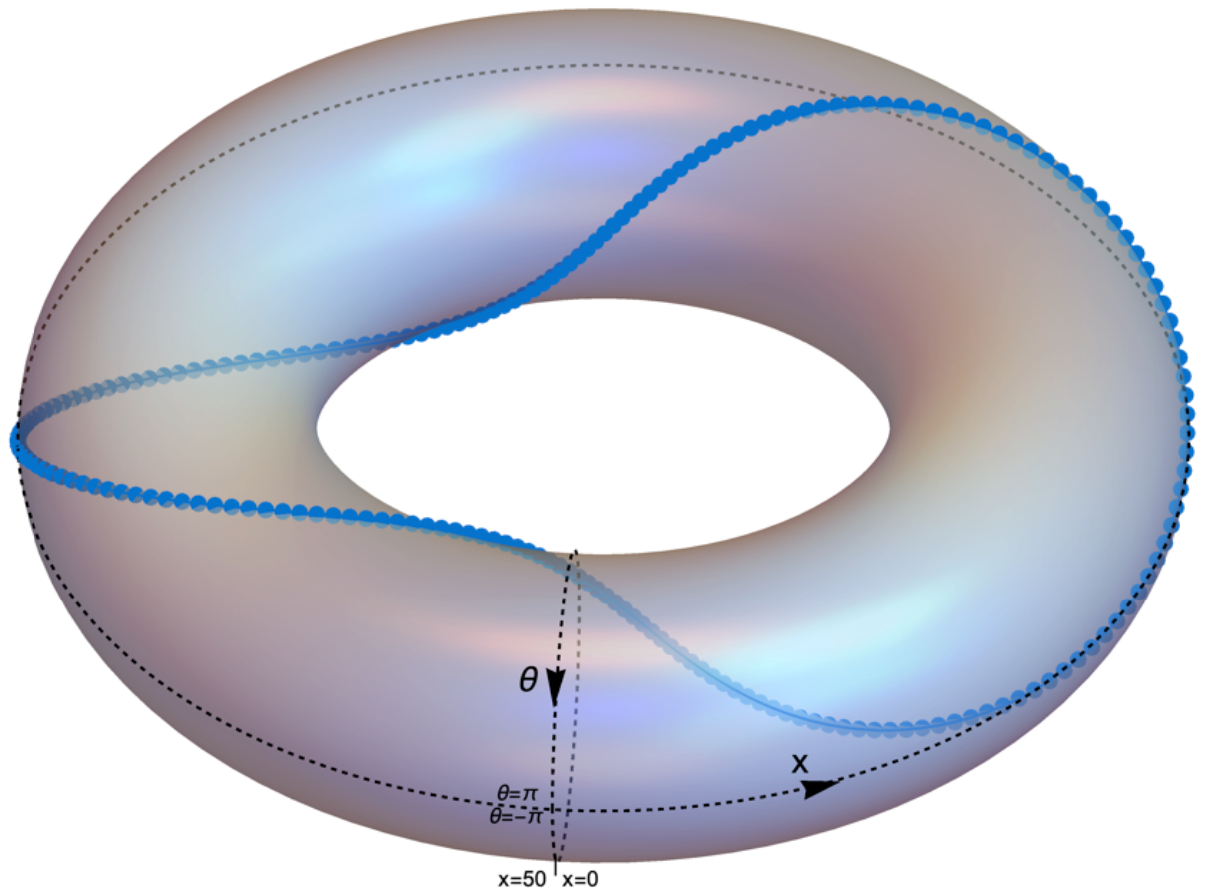}
\put(-200,0){(c)}~~~~~~~~~~~~~~~~~~~~~~~~
\includegraphics[trim=2.5cm 5.3cm 3cm 4cm, clip=true, angle=-90, scale=0.3]{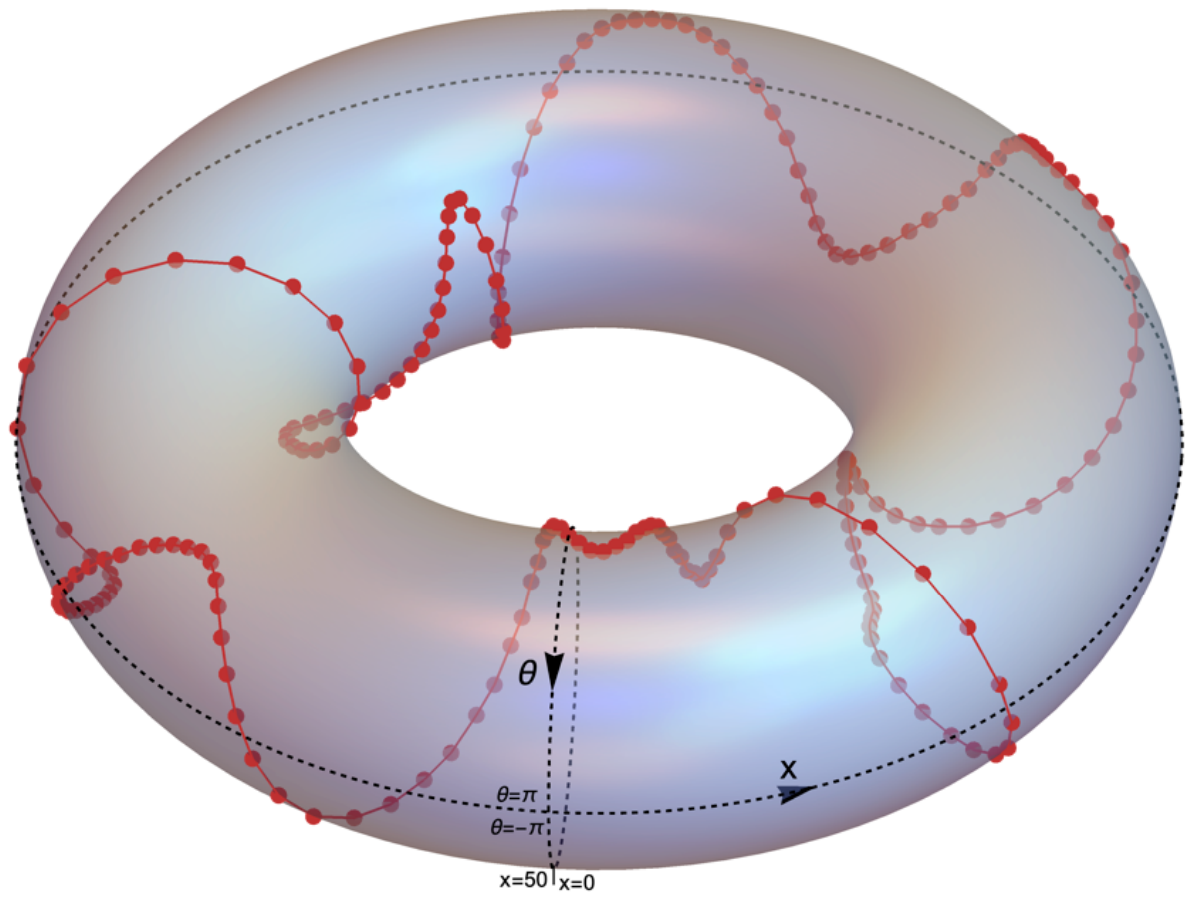}
\put(-200,0){(d)}
\caption{{\bf Time evolutions of the order parameter and the corresponding phases from initial temperature to final equilibrium state with the quench rate $\tau_Q=20$.} (a) and (b) show the temporal evolution of the average condensate of the order parameter and the phases $\theta$ at some specific times for holographic superfluid and superconductor system, respectively. The instant $t=80$ is the time when the condensate just arrives at its equilibrium value, however, the phases will finally relax at a later time, i.e., the final equilibrium time $t=1000$.  The phase in superfluid model in the final equilibrium state has constant gradient $\nabla\theta$ along the ring. On the contrary, the phase in superconductor at $t=1000$ is a smooth random function due to the gauge invariance.  The winding number at equilibrium state in superfluid is $W=-2$ (blue circles) and in superconductor is $W=+2$ (red circles). The dashed lines indicate spurious jumps of the phase at the edges $\theta=\pm\pi$. The stereographic views of the phases in (a) and (b) at time $t=1000$ are shown in (c) and (d), respectively. As we can see intuitively from (c), the phase wraps the torus counter-clockwise twice along the $x$-direction, while phase rotates clock-wisely in panel (d).}\label{p2}
\end{figure}

The winding number along a compact one-dimensional superfluid/superconductor ring can be defined as
\be\label{eqw}
W=\oint_\mathcal{C} \frac{d{\bf \theta}}{2\pi} =\oint_\mathcal{C} \frac{\nabla\theta}{2\pi} dx\in\mathbb{Z}
\ee
where $\mathcal{C}$ denotes the circumference of the ring and $\theta$ is the phase of the order parameter. In numerics, we set the length of the circumference to be $L=50$ (i.e., $x\in [0, 50]$). From KZM, topological defects -- winding numbers --  are expected to form as we quench the system across the phase transition point that breaks the $U(1)$ gauge symmetry along a ring \cite{Das:2011cx}. We quench the system from the initial temperature $T_i=1.4T_c$ ($t=-8.4$) to the final temperature $T_f=0.8T_c$, and then maintain the system at $T_f$ until it arrives at the final equilibrium state. 

Fig.\ref{p2} shows the time evolution of the average condensate of the order parameter and the corresponding phases $\theta$ at the times $t=80$ and $t=1000$. Panel (a) and panel (b) correspond to the holographic superfluid and holographic superconductor systems, respectively. In this figure, all the parameters for the superfluid and superconductor models are identical except for the boundary conditions of the gauge fields $A_x$ as we stressed above. Specifically, we set the Dirichlet boundary conditions for $A_x$ near $z\to0$ boundary, i.e, we set $a_x=0$ for the holographic superfluid model. Alternatively, we set the Neumann boundary conditions for $A_x$ near $z\to0$ boundary, i.e. to set $b_x=0$ for holographic superconductor model. These different boundary conditions will result in different phenomena for the two models, especially in the phase configurations which will investigate in the following. 

Time evolution of condensate of the order parameters are alike for holographic superfluid or superconductor systems, which can be seen from panel (a) and panel (b) in Fig.\ref{p2}. Both of the condensates firstly are very small in the early time, then at around $t=50$ the condensates develop some finite value and then increase rapidly. At around $t=80$ the condensates arrive at the early stage of the equilibrium state that the condensate will remain unchanged, but the phase of the order parameter will still undergo phase ordering dynamics. See the corresponding plots of the phases at $t=80$ and $t=1000$. For the holographic superfluid model in panel (a),  the phase at $t=80$ does not have the constant gradient along the ring, however, at the final equilibrium time $t=1000$ the phase will become constant gradient.  The dashed lines connecting the edge values of the phases are spurious jumps of the phases, i.e., the phases are actually smooth along the ring since the range of the phases are $\theta\in[-\pi,\pi]$. If one projects this plot at $t=1000$ to a two-torus with coordinates $(\theta, x)$, this smoothness of the phase will be obvious. Please see the panel (c) for the superfluid model in Fig.\ref{p2}. The winding numbers at final equilibrium state for the superfluid system is $W=-2$, and the phase wraps the torus counter-clockwisely twice along the $x$-direction. \footnote{In this paper, we define the winding number to be $W=\pm n$ ($n\geq 0$). ``$+$'' indicates the phase goes from $-\pi$ to $\pi$ while ``$-$'' is the opposite and $n$ represents the phase wrap it $n$ times along the $x$-direction. }.  

For the holographic superconductor system, the phase at the final equilibrium time $t=1000$ does not have the constant gradient, comparing to the phase of superfluid. The reason stems from the requirement of the minimum of the free energy and the different boundary conditions we imposed for these two models. We define the gauge invariant velocity as 
\be
{\bf u}=\nabla\theta-a_x.
\ee 
In the free energy there is a quadratic term of the gauge invariant velocity $(\nabla\theta-a_x)^2$ \cite{tinkham}. In order to have a minimum value of the free energy one usually needs to render $(\nabla\theta-a_x)^2=0$ in the equilibrium state. In the holographic superfluid model, there is no gauge fields on the boundary, i.e., $a_x=0$. Thus, usually one needs $\nabla\theta=0$ in the superfluid phase, which indicates a long-range coherence of the phase since phase is constant in space. However, in our model the holographic superfluid is living in a compact ring, which may result in non-trivial winding numbers $W\neq0$. Hence, from Eq.\eqref{eqw} we infer that for $W\neq0$, the phase $\theta$ cannot be a constant. A natural result is that at the final equilibrium state $\theta=(2\pi W/L) x+c_1$ for holographic superfluid model, where $c_1$ is a constant depending on the random seeds in the numerics. \footnote{The linearity of the phase in superfluid model can be simply derived from the requirement of the minimum of free energy. In the free energy there is a term proportional to $I=\oint dx (\nabla\theta)^2$. We can regard $\theta$ as a fundamental field, and then use the variational method to derive the Euler-Lagrangian equations. One needs to note that in the variation, the term $(\nabla\theta)\delta\theta\big|^L_0$ is vanishing due to the compact of the ring. Thus, the equation of motion is $\nabla^2\theta=0$, having the solution $\theta(x)=ax+c_1$. Imposing the constraint $\oint(\nabla\theta)dx/(2\pi)=W$, i.e. Eq.\eqref{eqw}, we arrive at the final solution $\theta(x)=(2\pi W/L)x+c_1$ where $c_1$ is a constant depending on the numerics.  } This is the reason that in the panel (a) of Fig.\ref{p2} the phase of superfluid has a constant gradient at time $t=1000$. 

However, for the holographic superconductors we impose the Neumann boundary condition for $A_x$, therefore there is a gauge field $a_x$ in the boundary field theory. Thus, the gauge invariant velocity ${\bf u}=\nabla\theta-a_x$ comes into play. Since we have imposed $b_x=\partial_zA_x|_{z\to0}=0$, the value of $a_x$ is obtained from the results of quenched dynamics. In the final equilibrium state, $a_x(x)$ is a smooth random function along the ring due to the statistical properties of quenched dynamics from the random seeds. Thus, in order to meet the requirement of the minimum free energy, $(\nabla\theta-a_x)^2$ should be zero, then $\nabla\theta=a_x$ in the final equilibrium state. Therefore, we see from panel (b) in Fig.\ref{p2} that at time $t=1000$ the phase of holographic superconductor does not have a constant phase, but rather it is a random smooth function along the ring. \footnote{$a_x$ is the quantity which is solved from the time evolutions of the equations of motions. Therefore, the randomness of the smooth function $a_x$ is due to the random seeds in the initial thermalization. } Its projection to the two-tours is shown in the panel (d), which vividly presents the randomness of the phase along the ring. It goes clock-wisely twice along the $x$-direction, thus having the winding number $W=+2$.


\section{Persistent superflow \& gauge invariant velocity}
\label{statistics}

This section will devote to discuss the relationships between the phase and the superflow velocity in the holographic superfluid and superconductor model. Moreover, the statistical distributions of the average value of $\langle a_x\rangle$ will also be investigated in holographic superconductor system. 

\begin{figure}[h]
\centering
\includegraphics[trim=1.5cm 7.cm 0.cm 7cm, clip=true, scale=0.42]{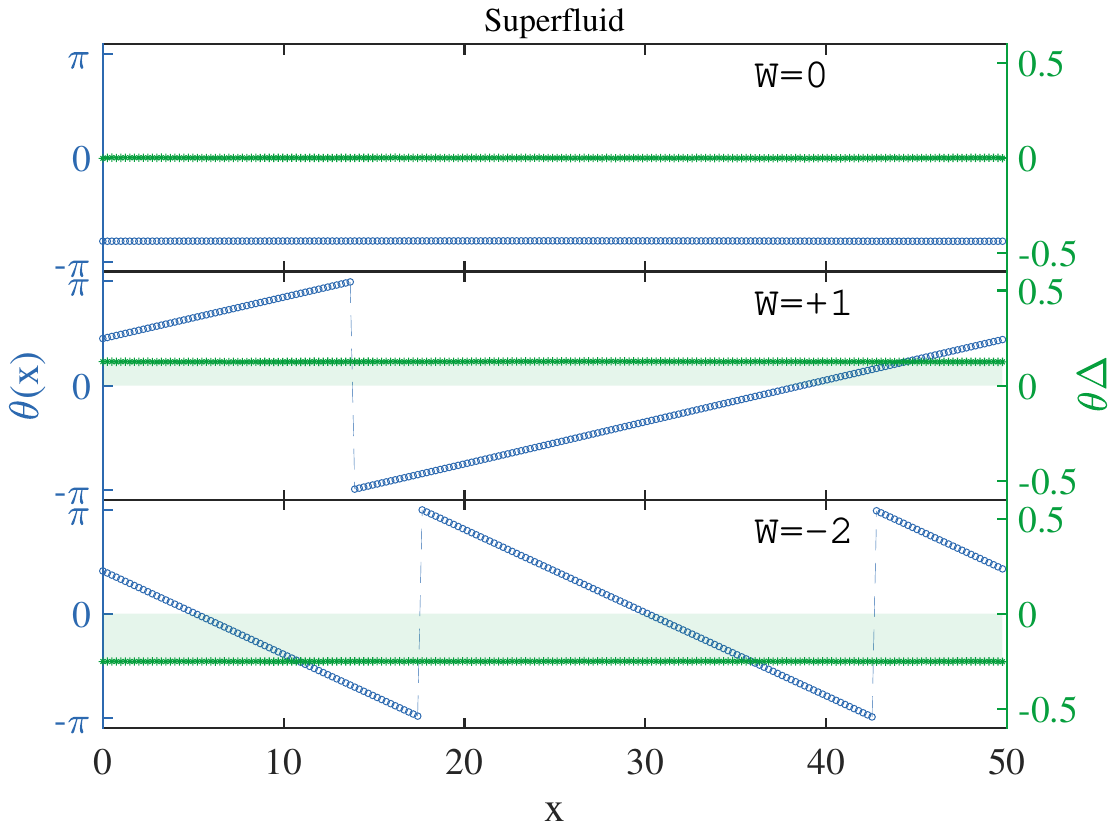}
\put(-245,170){(a)}~
\includegraphics[trim=1.5cm 7.3cm 2.2cm 7cm, clip=true, scale=0.42]{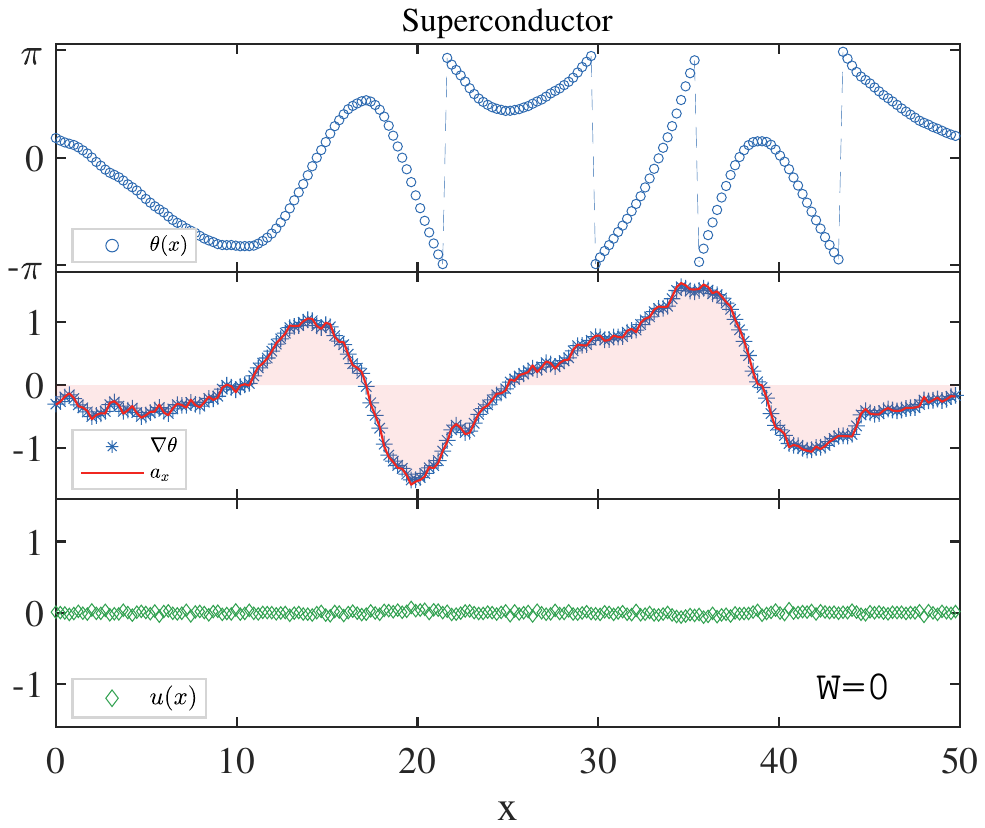}
\put(-215,170){(b)}~\\
\includegraphics[trim=1.5cm 7.3cm 0cm 7cm, clip=true, scale=0.42]{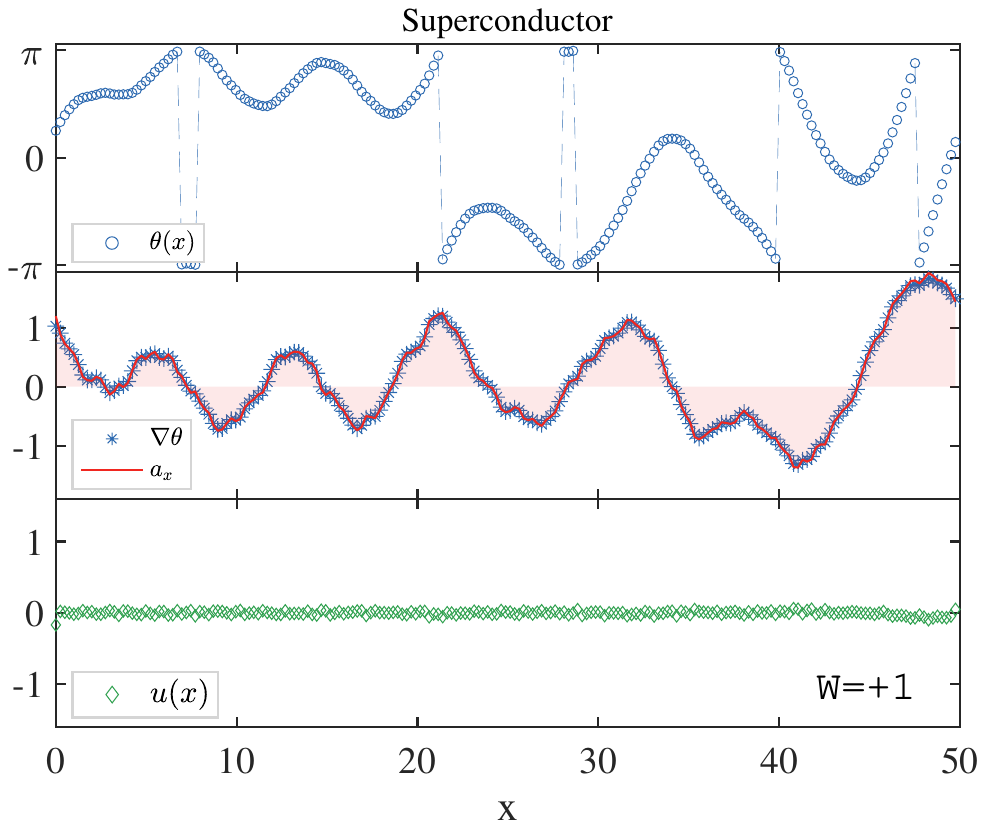}
\put(-245,170){(c)}~
\includegraphics[trim=1.5cm 7.3cm 2.2cm 7cm, clip=true, scale=0.42]{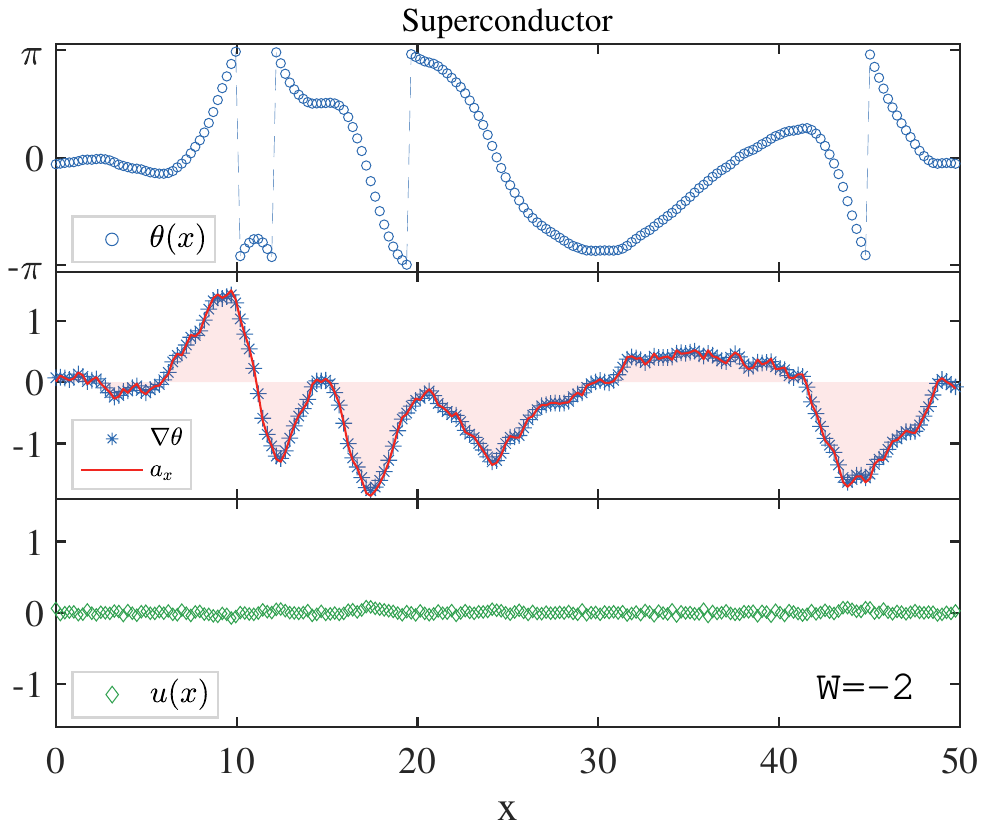}
\put(-215,170){(d)}~
\caption{{\bf Phase configurations and the gauge invariant velocity of the superflow with different winding numbers ($W=0, +1$ and $-2$) in the final equilibrium state.} (a) shows the phase $\theta(x)$ (blue circles) and the gradient of phase $\nabla\theta$ (green asterisks) in the holographic superfluid model. The areas of shaded green regions are proportional to the corresponding winding numbers. (b) - (d) show the phase configurations and the gauge invariant velocity $\bf{u}$ of the superflow in superconductor model with various winding numbers. The blue circles and asterisks represent the phase configurations $\theta (x)$ and gradient of phase $\nabla\theta(x)$, respectively. The red lines are the gauge field $a_x$ and the green diamonds stand for the gauge invariant velocity $\bf{u}$.  In addition, the shaded red regions are proportional to the corresponding winding numbers. In this figure, we set the quench rate $\tau_Q=20$.
}\label{scw}
\end{figure}

{\bf Superfluid:}
In the model of holographic superfluid, the Dirichlet boundary conditions for the gauge fields $a_x=0$ are imposed at the boundary $z\rightarrow 0$. Therefore, the superflow  velocity is exactly the gradient of the phase, i.e., $\bf{u} = \bf\nabla \theta$. We show the phase configurations of different winding numbers ($W=0, +1$ and $-2$) of superfluid (blue circles) and their corresponding velocity ${\bf u}=\nabla\theta$ (green asterisks) in the final equilibrium state ($t=1000$) with $\tau_Q=20$ in Fig.\ref{scw} (a). It is clear to see that $\bf{u}$ is constant at the final equilibrium state, implying a persistent superflow along the ring. The areas of the shaded green regions, that is, the integral of the velocity along the $x$-direction, are exactly equal to the corresponding winding numbers $W=\frac{1}{2\pi}\oint_\mathcal{C} {\bf u} dx$.

\begin{figure}[h]
\centering
\includegraphics[trim=1.cm 7.cm 1.5cm 7cm, clip=true, scale=0.5]{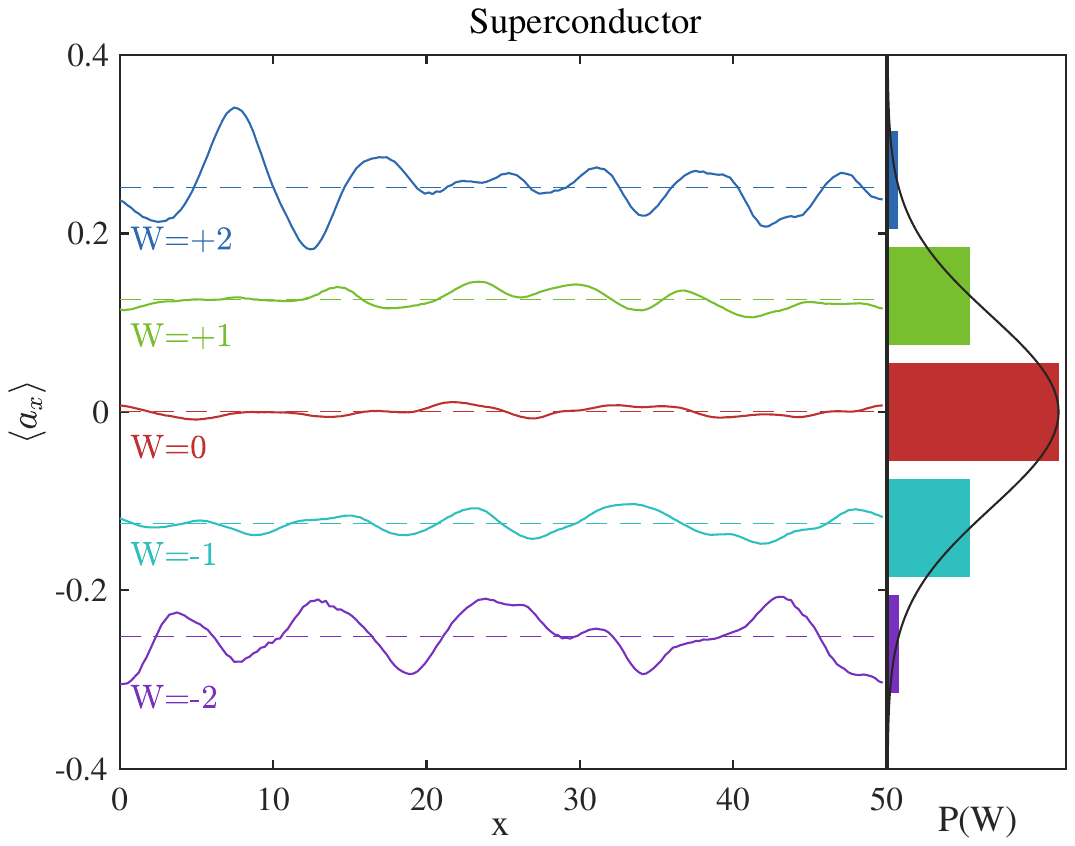}
\caption{{\bf Average values of gauge fields $\langle a_x(x)\rangle$ along the ring with different winding numbers and the corresponding statistics.} Left panel shows the average values of the gauge fields $\langle a_x(x)\rangle$ along the ring after $5000$ times of simulations, as the colored solid lines indicate. Each colored line corresponds to different winding numbers. The dashed lines correspond to the values $2\pi W/L$. Right panel exhibits the probability density $P(W)$ of the winding numbers. The histogram are consistent with the normal distribution (black line) with the mean $\langle W\rangle\approx0$ and variance $\sigma^2(W)\approx 0.8116$. }\label{wax}
\end{figure}

{\bf Superconductor:}
In contrast to holographic superfluid, we impose the Neumann boundary conditions for the gauge fields on the $z\rightarrow 0$ boundary in the holographic superconductors. From the discussions of the free energy in the preceding section, we know that in the final equilibrium state of the holographic superconductor we always have ${\bf u}=\nabla\theta-a_x=0$, as the green diamonds show in panel (b)-(d) in Fig.\ref{scw}. This can also be seen from panels (b)-(d) in Fig.\ref{scw} that $a_x$ (red lines) are always overlapping with $\nabla\theta$ (blue asterisks) for various winding numbers. Therefore, in the final equilibrium state the winding number can also be expressed as $W=\oint_\mathcal{C} d\theta/(2\pi)=\oint_\mathcal{C} a_xdx/(2\pi)$, implying that the areas of the red shaded regions in Fig.\ref{scw} (b)-(d) are proportional to their corresponding winding numbers.  As we have stressed in the preceding section, the values of gauge fields $a_x$ are obtained from the quenched dynamics in the holographic superconductor, which is different from the superfluid model that the value of $a_x$ is put by hand. Therefore, due to the local randomness of the gauge fields $a_x$, the final configurations of the gauge fields are the smooth random functions along the ring. Hence, the configurations of the phase $\theta(x)$ of the superconductor are also smooth random functions, as the panels (b)-(d) show in Fig.\ref{scw}. We further investigate the statistical properties of the gauge fields $a_x$ and the distributions of winding number $W$ in Fig.\ref{wax}. As we stated above, the gauge field $a_x$ in the final equilibrium state of a quenched superconductor ring admits a random smooth distribution along the ring, then its average value along the ring is expected to be $\langle a_x\rangle\approx 2\pi W/L$ according to the equality $W=\oint_\mathcal{C} a_xdx/(2\pi)$. By averaging the data along the ring over $5000$ times independent simulations, we get the averaged value $\langle a_x\rangle$ for each winding number as $\langle a_x(W=-2)\rangle\approx-0.2514$, $\langle a_x(W=-1)\rangle\approx-0.1257$, $\langle a_x(W=0)\rangle\approx-1.7326\times10^{-6}$, $\langle a_x(W=+1)\rangle\approx0.1257$ and $\langle a_x(W=+2)\rangle\approx 0.2511$. The numerical results satisfy the prediction very well. 

On the other hand, one can also average $a_x(x)$ at each position $x$ for each winding number $W$.  This relation is shown in the left panel of Fig.\ref{wax}, in which the average values of $\langle a_x(x)\rangle$ (solid colored lines) stay around the dashed lines $2\pi W/L$ along the ring. We see that as $W$ is smaller, the fluctuations of $\langle a_x\rangle$ around the dashed lines $2\pi W/L$ are more moderate. On the contrary, if $W$ is bigger, the fluctuations of $\langle a_x\rangle$ are more violent. This is because during these $5000$ times simulations, solutions with smaller winding numbers will appear more frequently. See the right panel of Fig.\ref{wax} of the probability density of the winding numbers during numerous simulations. The histogram of $W$ is consistent with the normal distribution with the mean $\langle W\rangle\approx0$ and the variance $\sigma^2(W)\approx 0.8116$. Therefore, there will be more solutions turn out for instance of $W=0$. Hence, the fluctuations of the average $\langle a_x(x)\rangle$ for $W=0$ will be milder since there are more samples to get this average value. On the contrary, the fluctuations of $\langle a_x\rangle$ for bigger $W$ will be stronger since there are fewer samples to be averaged. However, whatever $W$ is, $\langle a_x(x)\rangle$ will always fluctuate around the dashed line $2\pi W/L$ since their average is constrained by the gauge invariant quantity and the requirement of the minimum of free energy as we discussed above. { Similar normal distributions of the winding numbers were previously studied in \cite{Xia:2020cjl,Das:2011cx}.}



\section{Two-point correlation functions in the large $N$ limit}
\label{2point}

\begin{figure}[h]
\centering
\includegraphics[trim=1.cm 7.cm 2.2cm 7cm, clip=true, scale=0.43]{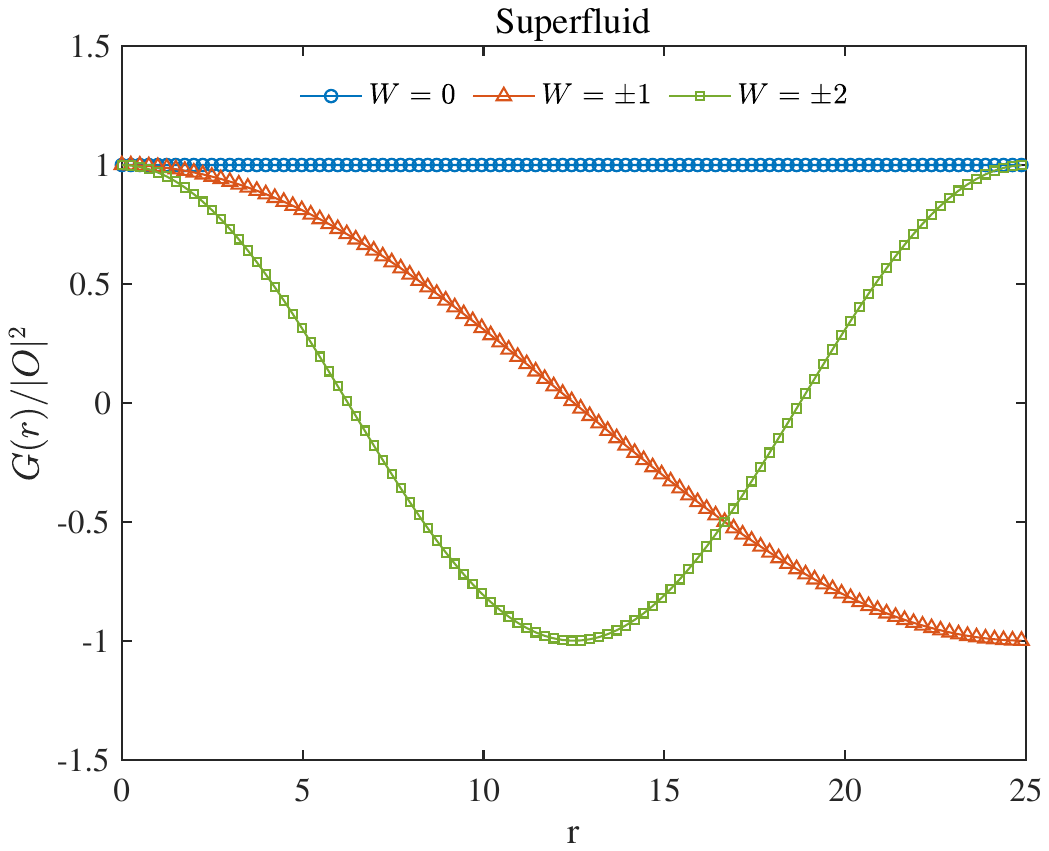}
\put(-215,180){(a)}~
\includegraphics[trim=1.cm 7.cm 2.2cm 7cm, clip=true, scale=0.43]{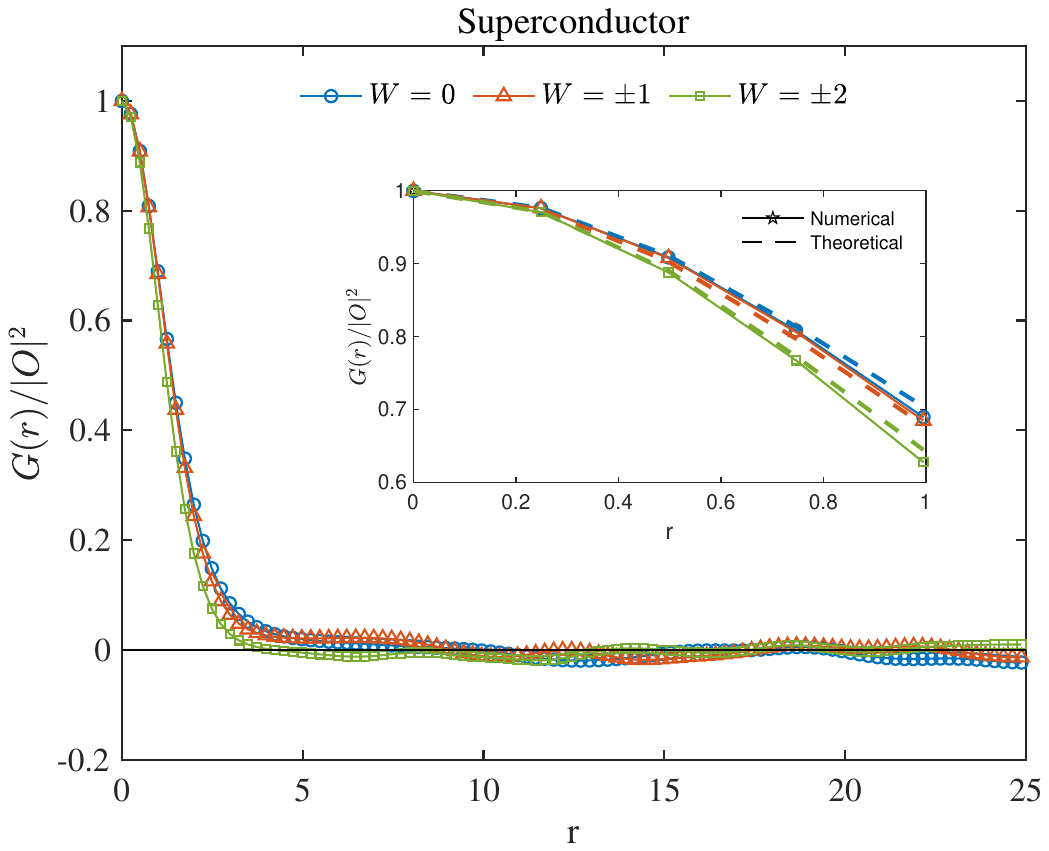}
\put(-215,180){(b)}~
\caption{{\bf Reduced two-point correlation functions $G(r)/|O|^2$ in holographic superfluid and holographic superconductor models for different winding numbers.} (a) Dependence of the correlation functions on $r$ in holographic superfluid model for different winding numbers. Numerical results of $G(r)$ match the analytic formula Eq.(\ref{oo}) very well. (b) Dependence of the correlation functions on $r$ in holographic superconductor model for different winding numbers. Their behaviors are dramatically distinct from those in panel (a). There are only short-range correlations for holographic superconductor. In the inset, we compare the numerical and theoretical correlations in short range, showing a consistent match.}\label{p4}
\end{figure}

In the following, we are going to study the two-point functions of the order parameter $G(x-y)$. In quantum field theory, $G(x-y)$ can be exactly computed as \cite{Zinn-Justin:2002ecy}, 
\be\label{2ptcorre}
G(x-y)\equiv\langle O(x)O(y)\rangle=\langle O(x)O(y)\rangle_c+\langle O(x)\rangle\langle O(y)\rangle. 
\ee
where on the r.h.s. the symbol $\langle\dots\rangle_c$ represents the connected functions while the second part  $\langle O(x)\rangle\langle O(y)\rangle$ is the disconnected functions. In the large $N$ limit of quantum field theory, the connected part is suppressed in the order of $N^{-2}$, thus the dominant part is the disconnected part, i.e., $\langle O(x)\rangle\langle O(y)\rangle$ makes a major contribution than $\langle O(x)O(y)\rangle_c$ \cite{Coleman:1985rnk,Zaanen:2015oix}. It is known that in the AdS/CFT correspondence, the boundary field theory is also in the large $N$ limit (here $N$ is the number of the colors in the AdS boundary field theory) \cite{Hartnoll:2009qx}, thus we can compute the two-point correlation function as $G(x-y)=\langle O(x)\rangle\langle O(y)\rangle+\mathcal{O}(N^{-2})$. \footnote{In AdS/CFT correspondence, people usually use the connected part $\langle O(x)O(y)\rangle_c$ to evaluate the two-point functions, by abandoning the disconnected part. This is because in that case the background is only a vacuum, i.e., $\langle O\rangle=0$, or the expectation value $\langle O\rangle$ is trivial, thus the disconnected part $\langle O(x)\rangle\langle O(y)\rangle$ has less information in space. However, as we will see later, due to the existence of windings of the phases, the disconnected part $\langle O(x)\rangle\langle O(y)\rangle$ is indeed non-trivial in space. Therefore, $\langle O(x)\rangle\langle O(y)\rangle$ is dominant in the two-point function $G(x-y)$. Thanks to the communications with J. Maldacena and E. Witten. }

In the 1D ring of the holographic superfluid/superconductor system, we will denote the two-point function as $G(r)\equiv\langle O(r)^\dagger O(0)\rangle$, in which we set one operator at the location $x=0$ and the other at $x=r$. We use the symbol $^\dagger$ because the operator is complex.  Thus, in the large $N$ limit, the correlation function $G(r)$ in the final equilibrium state can be reduced to 
\be
G(r)\underset{N\to\infty}{\sim}\langle O(r)^\dagger\rangle\langle O(0)\rangle.
\ee
In practice, we can put one of the condensation $\langle O\rangle$ at the origin $x=0$, and then multiplies the complex conjugate of the condensation value $\langle O^\dagger\rangle$ at the distance $r$ to this origin. There will be two such kinds of condensates having distance $r$ to the origin.  Then, we repeat this procedure again and again by putting all other condensates $\langle O\rangle$ as the origin, to evaluate the averaged values of $G(r)$. Since the ring is compact with the circumference $L$, the maximum distance $r$ to the origin should be $r_{\text{max}}=L/2$. The distance greater than $r_{\text{max}}$ is equivalent to the distance $L-r$. Thus, in figures we only plot the distance from $0$ to $r_{\text{max}}$.

{\bf Superfluid:}
For holographic superfluid model, the phase is linearly dependent on $x$ in the final equilibrium state that $\theta=(\nabla\theta)  x+\theta_c$,  where $\theta_c$ is an arbitrary value depending on the numerics.  Therefore, the average condensate of order parameter at a specific position $x=r_0$ in the equilibrium state can be expressed as
\be
\langle O(r_0)\rangle=\langle|O| e^{i\theta}\rangle=\langle|O| e^{i({(\nabla\theta)}r_0+\theta_c)}\rangle, ~~~
\langle O(r_0)^\dagger\rangle=\langle|O| e^{-i\theta}\rangle=\langle|O| e^{-i((\nabla\theta)r_0+\theta_c)}\rangle.
\ee
Since in the final equilibrium state the amplitude of the condensate is a constant along the ring, therefore, we can extract the amplitude $|O|$ out of the average value. Thus, we get the correlation function with distance $r$ from position $r_0$ as
\be
G(r)&\underset{N\to\infty}{\sim}&\frac{1}{2}\left(\langle O(r_0-r)^\dagger\rangle\langle O(r_0)\rangle+\langle O(r_0+r)^\dagger\rangle\langle O(r_0)\rangle\right)\nonumber\\
&=&\frac{1}{2}| O|^2\langle e^{i{(\nabla\theta)}r}+e^{-i{(\nabla\theta)}r}\rangle
=|O|^2 \langle\cos({(\nabla\theta)}r)\rangle
\ee
The final correlation function actually is independent of the referenced position $r_0$. Therefore, we can only write it as $G(r)$. From the Eq.\eqref{eqw} we can see that at the final equilibrium state $\nabla\theta=2\pi W/L$, thus the reduced correlation function can be written as,
 \be\label{oo}
 \frac{G(r)}{|O|^2} \underset{N\to\infty}{\sim}\left\langle\cos\left(\frac{2\pi  W}{L}r\right)\right\rangle.
\ee
This relation indicates that the two-point correlation functions are irrelevant to the amplitudes of the condensate $|O|$, but rather they depend on the phase correlations.  Fig.\ref{p4}(a) shows the numerical results of reduced correlation function $G(r)/|O|^2$ for holographic superfluid, in which the blue circles, red triangles and green squares represent the correlation functions with winding numbers $W=\{ 0, \pm1, \pm2\}$, respectively. These numerical results match the theoretical prediction Eq.(\ref{oo}) very well. In particular, we see that for $W=0$ the correlation function is a constant along the ring, which represents a perfect coherence between the phases. 
 
{\bf Superconductor:}
In contrast to superfluid model, the correlation function $G(r)/|O|^2$ for the order parameter of superconductor is very different. As we stressed previously, in the final equilibrium state the phase $\theta$ of the order parameter in superconductor is a random and smooth function. Thus, when $r$ is a little bit longer, it is expected that there is no correlations between the phases since it is destroyed by the randomness. The numerical results for the large distance correlations can be found in panel (b) of Fig.\ref{p4} that it vanishes at long distance, which satisfies our physical intuition. However, as $r\rightarrow r_0$ we can get a semi-analytical relation between the correlation and the distance as 
\be
G(r)&\underset{N\to\infty}{\sim}&\frac{1}{2}\left(\langle O(r_0-r)^\dagger\rangle\langle O(r_0)\rangle+\langle O(r_0+r)^\dagger\rangle\langle O(r_0)\rangle\right)\nonumber\\
&=&\frac{1}{2}| O|^2\left\langle e^{i[\theta (r_0)-\theta(r_0-r)]}+e^{-i[{\theta(r_0+r)-\theta(r_0)]}}\right\rangle\nonumber\\
&\underset{r\to r_0}{\sim}&\frac{1}{2}| O|^2\left\langle e^{i\nabla\theta (r_0)r}+e^{-i\nabla\theta(r_0) r}\right\rangle
=|O|^2 \left\langle\cos\left(\nabla\theta (r_0) r\right)\right\rangle,
\ee
where $\nabla\theta(r_0)$ is the gradient of phase at the location $x=r_0$. From the above discussion, we already know that in the final equilibrium state $\nabla\theta(r_0)= a_x(r_0)$ in superconductor, therefore, the reduced correlation function can be reduced as $G(r)/|O|^2\sim \left\langle \cos( a_x(r_0) r)\right\rangle$. Setting the specific point $r_0=0$, and Taylor expand it at $r = 0$, we get
\be\label{r0OO}
\frac{G(r)}{|O|^2} &\sim&1-\frac{ \langle a_x(0)^2\rangle}{2!}r^2+\mathcal{O}(r^3)
=1-\frac{ \sigma_{a_x}^2+\mu_{a_x}^2}{2}r^2+\mathcal{O}(r^3),
\ee
where $ \sigma_{a_x}^2$ and $\mu_{a_x}$ are the variances and the mean of $a_x$ at the specific point, respectively. These data can be readily obtained from the preceding section of the statistics. 

From Fig.\ref{p4}(b) we see that as $r$ increases away from $r=0$, the numerical results of the reduced correlation functions $G(r)/|O|^2$ drops to $0$ quickly. In the inset plot, the colored dashed lines have the relation $G(r\to 0) \propto 1-\frac{ \langle\sigma^2_i\rangle+\langle\mu_i^2\rangle}{2}r^2$, where $i$ represents different winding numbers and $\langle\dots\rangle$ denotes the average values at a specific location for independent simulations. The inset plot shows that the numerical results are consistent with the above semi-analytical equation Eq.\eqref{r0OO} as $r\to 0$. 

{Both two-point functions in holographic superfluids and superconductors are related to the phase correlations at different distances, since the absolute values of the operator, i.e., $\langle|O|\rangle$, are identical along the $x$-direction. The main differences between them is that for holographic superfluids, the phase correlations are the cosine function related to the winding number $W$ and the distance $r$, see Eq.\eqref{oo}. But for holographic superconductors, the phase correlations will decay rapidly at large distance $r$, refer to panel (b) of Fig.\ref{p4}.  }

\section{Concluding remarks}
\label{conclusion}
Taking advantage of the KZM, we dynamically realized the winding numbers of the order parameter in ring-shaped superfluid and superconductor model from gauge-gravity duality, by setting different boundary conditions of the spatial component of the gauge field near the AdS boundary. Due to the different boundary conditions of the two models, some of their results are quite different. In particular, at the final equilibrium state, we compared the configurations of the phases of the order parameters and the gauge invariant velocity of the two models. We found that the phase $\theta$ finally became `piecewise' straight lines in $x$-direction, implying a superflow with a constant velocity for superfluid model. However, for superconductor, since the gauge invariance constraint $\nabla\theta-a_x=0$ were always satisfied, the phase cannot eventually became a linear function. But rather, it became a smooth random function because of the random localities of the gauge fields. We further investigated the two-point correlation functions of the order parameter in superfluid and superconductor models. Their behaviors were dramatically different. Concretely, for superfluid model, the reduced correlation function $G(r)/|O|^2$ depended on the winding numbers $W$ as a cosine function, refer to the Eq.\eqref{oo}. On the contrary, the correlation function in superconductor model decayed rapidly and vanished in large distance, implying no correlations for long distance. As the distance was close to the origin, the correlation function exhibited as a decreasing function with a quadratic power in distance, which was consistent with numerical results.

\section*{Acknowledgements}
We appreciate the helpful discussions with J. Maldacena and E. Witten. This work was partially supported by the National Natural Science Foundation of China (Grants No. 11875095 and 12175008) and  partially supported by the Academic Excellence Foundation of BUAA for PhD Students.

{
\appendix

\section{Explicit forms of the equations of motions}
\label{app}
In the probe limit, the equations of motions for $\Psi$ and $A_\mu$ read,
\begin{eqnarray}\label{eomofwhole}
D_\mu D^\mu\Psi-m^2\Psi=0,~~~\nabla_\mu F^{\mu\nu}=i\left(\Psi^* D^\nu\Psi-\Psi{(D^\nu\Psi)^*}\right), 
\end{eqnarray}
From the ansatz $\Psi = \Psi(t,z,x), A_{t} = A_{t}(t,z,x), A_x=A_x(t,z,x)$ and $A_z =A_y= 0$, the explicit forms of the equations are, 
 \begin{eqnarray}
\label{eompsi}
\partial_t \partial_z \Phi - i A_t \partial_z \Phi - \frac12 [ i \partial_z A_t \Phi + f \partial_z^2 \Phi + f' \partial_z \Phi - z \Phi 
+ \partial_x^2 \Phi  - i (\partial_x A_x ) \Phi - A_x^2 \Phi - 2 i A_x \partial_x \Phi  ] = 0;~~~~~&
\\
\label{eom2}
\partial_t \partial_z A_t - \partial_x^2 A_t  - f \partial_z \partial_x A_x  + \partial_t \partial_x A_x 
+ 2 A_t |\Phi|^2 - i f (\Phi^* \partial_z \Phi - \Phi \partial_z \Phi^*) + i (\Phi^* \partial_t \Phi - \Phi \partial_t \Phi^*) = 0;~~~~~&
\\
\label{eom3}
\partial_t \partial_z A_x - \frac12 \left[ \partial_z (\partial_x A_t + f \partial_z A_x) - i (\Phi^* \partial_x \Phi - \Phi \partial_x \Phi^*) - 2 A_x |\Phi|^2 \right] = 0;~~~~~&
\\
\label{eom1}
\partial_z (\partial_x A_x  - \partial_z A_t) + i (\Phi^* \partial_z \Phi - \Phi \partial_z \Phi^*) = 0.~~~~~~&
\end{eqnarray}
where $\Phi=\Psi/z$.
The above four equations are not independent, in particular their L.H.S. satisfy the following constraint equation,
\begin{eqnarray}
-\frac{d}{dt}\text{Eq.\eqref{eom1}}-\frac{d}{dz}\text{Eq.\eqref{eom2}}+2\frac{d}{dx}\text{Eq.\eqref{eom3}}\equiv-2i\left(\text{Eq.\eqref{eompsi}}\times\Phi^*-c.c.\right)
\end{eqnarray}
where $c.c$ represents complex conjugate. Therefore, there are three independent equations for three fields,  $\Phi, A_t$ and $A_x$. Since $\Phi=\Psi/z$ is a complex field, this also means that there are four independent real fields for four independent real equations. It in turn implies that our choice of the gauge $A_z=A_y=0$ is viable for the setup of the system. 
}

\end{document}